# A generalization of Snoek's law to ferromagnetic films and composites

O. Acher*, S. Dubourg

CEA Le Ripault, BP16, 37260 Monts, France

**Abstract:**

The present paper establishes characteristics of the relative magnetic permeability spectrum $\mu(f)$ of magnetic materials at microwave frequencies. The integral $\int_0^\infty \mu''(f).f.df$ of the imaginary part of $\mu(f)$ multiplied with the frequency $f$ gives remarkable properties. A generalisation of Snoek's law consists in this quantity being bounded by the square of the saturation magnetization multiplied with a constant. While previous results have been obtained in the case of non-conductive materials, this work is a generalization to ferromagnetic materials and ferromagnetic-based composites with significant skin effect. The influence of truncating the summation to finite upper frequencies is investigated, and estimates associated to the finite summation are provided. It is established that, in practice, the integral does not depend on the damping model under consideration. Numerical experiments are performed in the exactly solvable case of ferromagnetic thin films with uniform magnetization, and these numerical experiments are found to confirm our theoretical results. Microwave permeability measurements on soft amorphous films are reported. The relation between $\int_0^{6GHz} \mu''(f).f.df$ and $4\pi M_s$ is verified experimentally, and some practical applications of the theoretical results are introduced. The integral can be used to determine the average magnetization orientation in materials with complex configurations of the magnetization, and furthermore to demonstrate the accuracy of microwave measurement systems. For certain applications, such as electromagnetic compatibility or radar absorbing materials, the relations established herein provide useful indications for the design of efficient materials, and simple figures of merit to compare the properties measured on various materials.

## I. INTRODUCTION

The microwave permeability $\mu$ of magnetic materials is a quantity of interest both with respect to applied and fundamental points of views. High frequency inductors,[1] magnetic recording write heads, broadband skin antennas,[2] microwave filters,[3,4] noise suppressors[5] and Radar Absorbing materials[6,7] require high broadband permeability levels at elevated frequencies. However, it has been known since the work of Snoek[8] that there exist tradeoffs between high permeability levels and operation at elevated frequencies, i.e. the higher the resonance frequency $F_0$, the lower is the low frequency permeability $\mu'_0$. In a bulk polycrystalline material, Snoek's law is written as

$$(\mu'_0 - 1)F_0 = \frac{2}{3}\gamma 4\pi M_s \qquad (1)$$

where $4\pi M_s$ is the saturation magnetization, and $\tilde{\gamma} = \gamma/2\pi \approx 3 MHz/Oe$ the gyromagnetic factor. The tradeoffs between the permeability level and the resonance frequency depend on the shape of the magnetic domains or particles[9,10] For soft thin films with uniform uniaxial in-plane anisotropies, the equation becomes

$$(\mu'_0 - 1)F_0^2 = (\tilde{\gamma} 4\pi M_s)^2. \qquad (2)$$

These relations are easily established from the gyromagnetic permeability of a saturated ellipsoid. However, they become invalid for heterogeneous magnetic materials, or in the case of composites. In addition, they provide no clue to the linewidth of the permeability.

Recently[11,12] another expression of the tradeoffs between permeability levels and frequency has been established[13]

$$\int_0^\infty \mu''(f).f.df = k_A \frac{\pi}{2}(\tilde{\gamma} 4\pi M_s)^2, \qquad (3)$$

Here, $k_A$ is a dimensionless factor associated with the distribution of the orientation of the magnetization in the sample. This sum law finds its root in the causality principle, associated with the fundamental equation of gyromagnetic motion. For uniform soft thin films, $k_A=1$. For bulk sintered ferrites, $k_A=1/3$. For isotropic composite materials with a volume fraction $\tau$ of a magnetic filler, $k_A \leq \tau/3$. In any case, $k_A \leq 1$. The ratio $k_A$ is easily determined from experimental data, and can be used to quantify the quality of thin films for microwave applications[13] and to guide their design.[14] Eq. (3) has also been found useful as an indication for the conception of microwave absorbers.[15] In this case, $\mu''$ is a quantity of direct interest.

Eq. (3) has a different form as compared to the original Snoek's law given by Eq. (1) and its extension to thin films given by Eq. (2). However, strong connections exist between these identities. According to Eq (3), the higher the resonance frequency (generally corresponding to the peak of $\mu''$), the lower is the bandwidth multiplied by the maximum $\mu''$ levels.

For materials with a permeability that coincides with the gyromagnetic permeability of a saturated ellipsoid, Snoek's law in its discrete form (i.e. Eqs. (1), (2)) may be a more straightforward expression of the balance between high permeability levels and operation at high frequencies as opposed to Eq. (3).[16] However, in many cases, Snoek's law does not apply in its discrete form whereas Eq. (3) remains valid. As a consequence, Eq. (3) can be considered as a generalization of Snoek's law. Hexagonal ferrites used in microwave applications are not soft materials in the sense that their out-of-plane anisotropy field is comparable or larger than the saturation magnetization. Thus, Eq. (3) does not apply to hexagonal ferrites, but a more general integral relation has been proposed and verified experimentally.[17]

The purpose of this paper is to provide significant extensions for Eq. (3). In its original derivation,[12] the effect of the conductivity on the permeability has been neglected. This is a significant limitation, since skin effect due to finite conductivity may substantially affect the permeability of ferromagnetic materials. An important result reported in this paper is that the integral of $\mu''(f).f$ is hardly affected by moderate skin effect, and slightly decreases when the skin effect becomes larger. Another limitation of Eq. (3) concerns the model used to describe the magnetic damping. This identity was established assuming a damping mechanism used in the equations of Bloch-Bloembergen. The present study demonstrates that it also holds true when the Gilbert description of the magnetization relaxation is employed.

When the integral in the left-hand side of Eq. (3) is determined from experimental permeability measurements, it has to be truncated to a finite upper frequency within the measurement range. This paper provides estimates of upper integration frequencies that can be used with negligible error, as well as simple estimates for the truncation error.

The paper is organized as follows. In part II, the theoretical approach is outlined and general results are presented. Analytical details are given in the appendix. In part III, results are formulated for particular cases, namely thin films, multilayers, and composite materials. In part IV, the relevancy of the approximations is verified through numerical experiments on an exactly solvable case. In part V, our theoretical findings are confronted with experimental values of permeabilities measured on thin films. Ultimately, the potential applications of our findings are discussed in part VI.

## II. THEORETICAL APPROACH

### A. Permeability of an ellipsoid with uniform magnetization

The magnetic susceptibility tensor of an ellipsoid with a uniform magnetization is well known.[18] Several models have been proposed for the damping mechanism. The Landau-Lifschitz-Gilbert expression figuring the dimensionless damping parameter $\alpha$ is one of the most popular. The full expression of $\mu_G$ is given in the appendix, see Eq. (A1). It depends on the saturation magnetization $4\pi M_s$ of the material, on the demagnetizing coefficients of the ellipsoid $N_x$, $N_y$, $N_z$, and on the resonance frequency $F_0$. It is convenient to introduce the quantity $F_M = \gamma 4\pi M_s$ which has the dimension of frequency.

### B. Influence of skin effect on the permeability

The permeability of a conductive inclusion depends on its conductivity, shape and dimensions, and of course on the permeability of the constitutive material. It has been derived within many independent studies, and for numerous shapes. The permeability of the inclusion can be written as[6,19,20,21,22]

$$\mu = \mu_G . A(ka), \quad (4)$$

where $\mu_G$ is the intrinsic permeability of the material, $k$ the wavevector associated to the microwave excitation inside the inclusion, and $a$ the radius of the inclusion (in the case of a sphere and a cylinder) or its half thickness (in the case of a plate). $k$ depends implicitly on the permeability $\mu_G$ and the conductivity $\sigma$ of the material. The expression of $A(ka)$ for various inclusion shapes is presented in Fig. 1. Though the expressions may appear dissimilar at first sight, their first order development in $ka$ has the same form

$$A(ka) = 1 + \frac{(k.a)^2}{p} + higher\ order\ terms \quad (5)$$

Here, $p$ is a number that depends on the shape of the inclusion. It can be seen in Fig. 1 that $p$ is larger for a sphere ($p=10$) than for a plate ($p=3$), and for a cylinder, its value is somewhere in between. This suggests that Eq. (5) can be extended to a variety of regular shapes, and that it is fairly general.

### C. Derivation of the integral bound

As a consequence of the causality principle, the permeability $\mu$ function of the complex variable $f$ is analytic in the lower half of the $f$-plane.[23] The Cauchy theorem is applied to the quantity $f.\mu(f)$ on a closed contour consisting in the $[-F, +F]$ segment of the real axis and the half circle $C^-$ defined by $F.\exp(j\theta)$, $\theta$ ranging from 0 to $\pi$. This yields

$$\int_{-F}^{F} \mu(f).f.df + \int_{C^-} \mu(f).f.df = 0. \quad (6)$$

The first term in Eq. (6) can be transformed using the general properties $\mu(-\bar{f}) = \bar{\mu}(f)$, where the bar corresponds to the conjugate. This yields an integral of $\mu''(f).f$. The second term can be transformed into an integral on the angular coordinate $\theta$ of the semicircle $C^-$. This is relatively easy to calculate provided that $F$ is large enough, but not too large, and is a result of good approximations of $\mu$ being available at high frequencies $F$. Detailed calculations are shown in the appendix. One finds

$$\int_0^F \mu''(f).f.df \approx \frac{\pi}{2} N_y F_M^2 .[1-t-s\pm e], \quad (7)$$

where $t$ and $s$ are small positive numbers, and $t$ corresponds to the finite truncation whereas $s$ is related to skin effect. The term $e$ is the error induced by the measurement and $\Delta\mu$ the uncertainties with respect to $\mu$.

$$t = \frac{2}{\pi}\alpha(2N_x + N_y)\frac{F_M}{F} \quad (8)$$

$$s = \frac{4\mu_0 a^2 \sigma}{p} N_y \frac{F_M^2}{F} \quad (9)$$

$$|e| \leq \frac{\Delta\mu}{\pi N_y}\left(\frac{F}{F_M}\right)^2. \quad (10)$$

Here, $\mu_0$ is the permeability of vacuum. The validity of Eqs. (7) through (10) requires that the upper summation frequency $F$ verifies the conditions in the first 4 lines of Table I. For most microwave magnetic materials, the permeability can be measured in the appropriate spectral range using conventional measurement systems, as will be evidenced in sections IV and V. Table I also indicates under which conditions the corrective terms $t$, $s$ or $e$ are negligible. In the case where they are small but not negligible, they can be determined from the analytical expression above.

### D. A generalization to magnetic materials with complex magnetization states and to composites

Let us now deal with materials in demagnetized states. For this purpose, we consider magnetic matter constituted of a collection of magnetic domains with various shapes. Each magnetic domain can be described as a saturated ellipsoid, with possibly differing demagnetizing coefficients and internal fields. Let us also allow some non-magnetic matter. All the inclusions are supposed to be much smaller than the wavelength. The permeability of this complex matter can be determined through an appropriate homogenization law. In practice, the difficulty is that the homogenization law depends not only on the permeability of each domain, but also on the details of their geometries and arrangement. In cases where only the permeabilities of the constituents are known, but not the exact topology, it is not possible to precisely determine the permeability $\mu_{eff}$ of the homogenized medium. It is nevertheless possible to know certain bounds on the complex values of $\mu_{eff}$. Different sets of bounds, known as Wiener, Hashin-Shtrikman and Milton-Bergman bounds,[24,25,26,27] have been derived, depending on the partial information available. At very high frequencies, the permeability $\mu_i(F)$ of each constituent is close to unity. In this case, all bounds converge to a single value of $\mu_{eff}$ that is independent of the composite topology:[12,28]

$$\mu_{eff}(F) \approx \sum_i \tau_i .\mu_i(F) = \langle \mu_i(F) \rangle, \quad (11)$$

where $\tau_i$ designates the volume fraction of each domain labelled $i$, and $\langle \ \rangle$ corresponds to a volume average. Eq. (11) is also valid for any complex frequency on the semicircle $C^-$. Using this result in Eq. (6), one finds:

$$\int_0^F \mu_{eff}''(f).f.df = \sum_i \tau_i \int_0^F \mu_i''(f).f.df . \qquad (12)$$

This identity is similar to Eq. (20.6) in ref. [24], and constitutes a very important result: the integral associated to the composite medium is simply the linear average of the integral associated to each constituent.

### E. Isotropic composites made of ferromagnetic loads in a dielectric matrix

In an isotropic material, $\mu_{eff} = (\mu_x + \mu_y + \mu_z)/3$. The integral $(\mu_{x,i} + \mu_{y,i} + \mu_{z,i})/3$ can be easily calculated for each ellipsoid, and then averaged on the whole sample by using Eq. (12). The volume fraction of magnetic particles is denoted $\tau$, and the average of the demagnetizing coefficients of the domains along their magnetization is denoted $\langle N_{//} \rangle$. $\langle N_{//} \rangle$ is expected to be small as compared to unity, since the magnetization tends to be aligned in the elongation direction in soft materials. The demagnetizing coefficient in the elongation direction of an ellipsoid is close to zero for large aspect ratios,[18] and it will therefore be treated as a first order correction. We obtain from Eqs. (7) and (12):

$$\int_0^F \mu''(f).f.df \approx \frac{\pi}{6} \tau (\gamma 4\pi M_s)^2 \left(1 - \langle N_{//} \rangle - \langle t \rangle - \langle s \rangle \pm e \right) \qquad (13)$$

with

$$\langle t \rangle = \frac{2}{\pi} \langle \alpha \rangle (1+\eta) \frac{F_M}{F} \qquad (14)$$

where $\eta = 2\langle N_x.N_z \rangle$ is the average of the product of the demagnetizing coefficients normal to the magnetization. This term is comprised between 0 and ½.

It is important to note that $s$ does not depend on the internal fields, which suggests that for a multi-domain particle with half radius $a$ and shape parameter $p$, Eq. (9) is still the appropriate expression for $s$, provided that an averaging on $N_y$ is performed. In practical cases, there may be a significant dispersion in radius $a$ within the magnetic filler, while its conductivity $\sigma$ remains constant. It thus follows that

$$\langle s \rangle = \frac{4\mu_0 \sigma \langle a^2 \rangle}{p} (1-\eta) \frac{(F_M)^2}{F} . \qquad (15)$$

The absolute error due to measurement uncertainties is bounded by $\frac{\Delta \mu}{2} F^2$, and a majoration of the relative error $e$ can be written as :

$$|e| \leq \frac{3}{\pi} \frac{\Delta \mu}{\tau} \left( \frac{F}{F_M} \right)^2 . \qquad (16)$$

### III. RESULTS

Eqs. (13) to (15) are very general and only valid when the upper integration frequency $F$ is chosen in the proper range as summarized in the four first lines in Table I. It is useful to rewrite these equations for a few cases of particular interest, in order to provide ready-to-use expressions.

## A. Application to soft thin films with uniform magnetization

In the case of a uniaxial thin film magnetized along the $z$ direction with in-plane orientation, the demagnetizing coefficient normal to the film plane $N_y$ is unity. The hard axis permeability (along $x$) has the following properties:

$$\int_0^F \mu''(f).f.df \approx \frac{\pi}{2}(\gamma 4\pi M_s)^2.[1-t-s\pm e], \quad (17)$$

with

$$t = \frac{2}{\pi}\alpha \frac{\gamma 4\pi M_s}{F}, \quad (18)$$

$$s = \frac{4\mu_0 a^2 \sigma}{3} \cdot \frac{(\gamma 4\pi M_s)^2}{F}, \quad (19)$$

$$|e| \leq \frac{\Delta\mu}{\pi}\left(\frac{F}{\gamma 4\pi M_s}\right)^2, \quad (20)$$

where $2a$ it the thickness of the film. When measurement errors are neglected, it follows in a straightforward manner that :

$$\int_0^F \mu''(f).f.df \leq \frac{\pi}{2}(\gamma 4\pi M_s)^2. \quad (21)$$

This generalizes previous results[11,12] to the case where a significant skin effect is present, provided that the summation is performed up to a reasonably high frequency. This demonstrates that skin effect cannot increase the factor of merit[13] $k_A$ defined in Eq. (3). Indeed, the presence of skin effect tends to decrease $k_A$, but only as a first order correction. This appears in Eq. (17) where -$s$ is a negative corrective term associated to skin effect.

The integral in Eqs. (17) and (21) can be easily evaluated from experimental results. The upper integration frequency $F$ should be chosen significantly larger than the resonance frequency, but smaller than $\gamma 4\pi M_s$.

The integral is expressed in Hz$^2$, and yields a number that is expected to have a limited significance to people working on magnetic materials. It is not easy to know whether the integral is large or small as compared to values observed on other microwave materials if it is expressed in Hz$^2$. One way to transform the integral into an easily interpreted number is to normalize the integral by $\frac{\pi}{2}(\gamma 4\pi M_s)^2$. This quantity has been introduced in Eq. (3) as the dimensionless parameter $k_A$. As aforementioned, this parameter is convenient to use, but does however require a prior determination of the saturation magnetization. Another solution is to turn the integral into a quantity that has the dimension of magnetization. This can be carried out very easily by taking the square root of the integral, and multiplying it by an appropriate constant. Let us define $M_\mu(F)$ by:

$$M_\mu(F) = \frac{1}{\gamma}\sqrt{\frac{2}{\pi}\int_0^F \mu''(f).f.df} \quad (22)$$

This quantity can be deduced from the permeability measurements with no other knowledge of the magnetic material. For a film with perfect orientation and limited skin effect, this quantity represents the saturation magnetization : $M_\mu(F)=4\pi M_s$. Thus, $M_\mu(F)$ should be a number with an intuitive signification to people involved in magnetic materials. In the general case:

$$M_\mu(F) \leq 4\pi M_s \tag{23}$$

Since the quantity $M_\mu(F)$ has the dimension of a magnetization and is obtained purely from dynamic measurements, it may be termed "efficient dynamic magnetization". It is also possible to estimate the saturation magnetization from the integral of the imaginary part of the permeability

$$4\pi M_s = M_\mu(F)\left(1 + \frac{t+s}{2} \pm \frac{e}{2}\right) \tag{24}$$

where $s$, $t$ and $e$ can be easily computed from Eqs. (18)-(20).

### B. Application to soft films with magnetization dispersion and multilayers

Let us consider a multilayer made of thin films with in-plane magnetization, but with the possibility of non-uniformities along its thickness. Such non-uniformities may arise from differences in anisotropy between the layers,[29] from antiferromagnetic coupling,[30] or from exchange coupling, provided that these fields are much smaller than the saturation magnetization. Non-uniformities may also arise from unwanted phenomena[31], or interfacial anisotropies. Let us also allow a certain non-uniformity within the film plane, provided that the demagnetization coefficient normal to the film plane remains close to unity. The angle between the $x$ direction in the film plane and the magnetization is denoted $\phi$. Then, neglecting measurement errors, the permeability $\mu_x$ along $x$ has the following properties:

$$\int_0^F \mu_x''(f).f.df = \frac{\pi}{2}\left\langle \sin^2\phi.(\gamma 4\pi M_s)^2 \right\rangle [1 - t - s]. \tag{25}$$

It follows that the integral of $\mu''$ can be a very useful tool to obtain insights regarding the orientation of the magnetization within multilayers. In the case where $4\pi M_s$ is uniform within the film but the orientation fluctuates, it is possible to get information on the orientation from permeability measurements along the $x$ and $z$ directions in the film plane:

$$\frac{\langle \sin^2\phi \rangle}{\langle \cos^2\phi \rangle} = \frac{\int_0^F \mu''_x(f).f.df}{\int_0^F \mu''_z(f).f.df}. \tag{26}$$

Though the permeability spectra along different directions are expected to be significantly influenced by the detailed topology of the magnetization dispersion, the ratio of integrals provides simple numerical indications on the average values of $\sin^2\phi$ and $\cos^2\phi$. This ratio has already been used in order to assess the effect of field annealing on the orientation of the magnetization.[32] The result presented here extends the validity of the method to thicker films.

Multilayers have also been designed to provide both magnetic softness and high saturation magnetization, by alternating materials with different saturation magnetization but with the same uniaxial in-plane orientation.[29] In this case

$$\int_0^F \mu_x''(f).f.df = \frac{\pi}{2}\sum_n \tau_n (\gamma 4\pi M_{s,n})^2 [1 - \langle t \rangle - \langle s \rangle \pm e]. \tag{27}$$

where $\tau_n$ is the volume fraction of material with a saturation magnetization of $4\pi M_{s,n}$. The corrective terms can be easily expressed from Eqs. (8)-(10), or neglected in the case where the summation is performed up to a sufficiently high frequency.

**C. Application to isotropic composites constituted of a magnetic load in a dielectric matrix**

Microwave composites made of a magnetic spherical powder dispersed in a dielectric matrix are widely used as microwave materials[33,34,35] and magnetic absorbers.[6,7,15] In most cases, the radius of the particles is significant as compared to the skin depth. Eq. (13) establishes that

$$\int_0^F \mu''(f).f.df \leq \frac{\pi}{6}\tau(\gamma 4\pi M_s)^2 . \qquad (28)$$

This result has already been theoretically obtained for composites made up of insulating materials such as ferrite powders, as well as verified experimentally.[12] The present work further established that this result remained largely unaffected by finite frequency summation, and by skin effect.

Eq. (15) (where $p=10$ is used for a sphere) provides useful guidelines for choosing an appropriate granulometry $<a^2>$ of the particles in order to obtain a small or negligible loss of dynamic permeability in the range of interest. The quantity

$$k_{A,3D} = \int_0^F \mu''(f).f.df \Big/ \left[\frac{\pi}{6}\tau(\gamma 4\pi M_s)^2\right] \qquad (29)$$

is a dimensionless figure of merit for isotropic composites. The closer it is to unity, the better is the material.

## IV. NUMERICAL VALIDATION

The aim of this numerical validation was to establish with confidence the exactitude of the estimates of the corrective terms $-t, -s$. Confirmation was also desired that the terms $+S'$ and $+g$ from Eq. (A17) could be neglected. The validation was performed on conducting thin magnetic films with uniform in-plane magnetization. This case was numerically simple, and also quite representative of typical measurements on soft ferromagnetic films. The numerical experiments were carried out with a precision that could not be attained in real experiments.

The thin film was described by the following parameters: $4\pi M_s=10$ kG, $H_k=16$ Oe, $\gamma=3$ GHz/kOe, $\alpha=2\%$, $\sigma=[130\,\mu\Omega.\text{cm}]^{-1}$, $N_y=1$, which are common values for thin soft ferromagnetic films. The resonance frequency was $F_0=1.2$ GHz. The relative permeability could be calculated using the Landau-Lifschitz Gilbert equation (Eq. (A1)) and the skin effect correction (Eq. (4), with the expression of $A(ka)$ for a plate according to Fig. 1). The imaginary part of the permeability is presented in Fig. 2 for thicknesses $2a$ ranging from 0.1 μm up to 2 μm. The broadening observed on the spectra of the 1 and 2 μm thick samples was due to skin effect. The numerical integration of $\mu''(f).f$ could be easily performed using these spectra.

Table II provides relevant lower and upper frequency bounds on the upper integration frequency in the case of the 2 μm thick film. It appears that $F=6$ GHz is in the adequate range. The values of $\int_0^{6GHz}\mu''(f).f.df$ obtained both numerically and analytically are displayed, and the numerical integration was performed on the spectra shown in Fig. 2. The analytical estimate for the integral was obtained using Eq. (17), and the corrective terms were evaluated, both numerically, and analytically using Eq. (18)-(20) and (A18), (A21). It appeared that the corrections $g$ and $s'$ were extremely small, and they could thus be neglected in the following. The effect of the truncation $-t$ was to underestimate the integral by approximately 6%, and our analytical estimate agreed well with the numerical determination. The skin effect correction $-s$ was -19% according to our analytical estimates, which was close to the -23% determined in our numerical experiment. In conclusion, the value of the integral predicted using our model displayed a deviation of less than 5% from the experimental value, for $F$=6GHz.

The influence of the upper integration frequency $F$ was also investigated in this numerical experiment. The "efficient dynamic magnetization" $M_\mu(F)$ defined by Eq. (22) is represented in Fig. 3. Though the expression of $M_\mu(F)$ may appear less friendly than the integral $\int_0^F \mu''(f).f.df$, it is very easy to calculate. Moreover, its value has an intuitive interpretation and can be compared directly to the saturation magnetization of the material. Numerical results were obtained by numerical integration of the permeability calculated using Eqs. (A1) and (4), whereas analytical results were established from the saturation magnetization using Eq. (24) and the values of $s$, $t$ and $e$ as computed from Eqs. (18)-(20). The analytical and numerical results were in excellent agreement, thus validating our results. For the 0.1 µm thick layer, the skin effect was negligible, and as illustrated in the graph, the truncation effects decreased when $F$ was increased.

The effect of measurement uncertainties was also explored. At frequencies much higher than the resonance frequency, $\mu''$ was weak, but could be affected by significant measurement uncertainties. On most permeability measurement systems for thin films, errors decrease when the thickness of the material increases as a result of a larger amount of magnetic material in the cell. For the 0.1 µm thick sample, a typical error $\Delta\mu$=10 was assumed, for the 1 µm thick film, $\Delta\mu$=2, whereas $\Delta\mu$=1 for the 2 µm film. The error bar associated with the integration is represented in Fig. 3. It can be seen that for the thinner film, the upper integration bound should not exceed $F$=6 GHz to a great extent, in order for $M_\mu(F)$ not to be significantly affected by the measurement uncertainties. On the thicker films, higher upper integration frequencies were possible with a satisfactory precision, but they may not be necessary. It was remarkable to see that the integral provided the value of the saturation magnetization within 10% if $F$>3.5 GHz for the 1 µm thick film, and if $F$>8 GHz for the 2 µm thick film. Behind the profound changes in the magnetic losses due to skin effect that can be evidenced in Fig. 2, it appears that the integral quantity $\int \mu''(f).f.df$ was nearly an invariant.

## V. EXPERIMENTAL VALIDATION

Amorphous CoZr thin films were sputter-deposited onto continuously transported 12 µm polyethylene teraphtalate substrates. The base pressure inside the chamber before deposition was less than $10^{-6}$ mbar, and during the process, the Ar pressure was fixed at $5.10^{-3}$ mbar. The residual magnetron field induced a uniaxial anisotropy parallel to the transportation direction. Four samples with various thicknesses, i.e. 0.3, 1.3, 1.7 and 2.1 µm, were fabricated. The saturation magnetization $4\pi M_s$ was measured using a Vibrating Sample Magnetometer, and was found to be 11.3 kG ± 0.5 kG. The permeability was determined using a thin film permeameter described elsewhere.[36] The typical error $\Delta\mu$ was estimated to approximately 20 for the thinnest film, and to 2 for the thicker ones.

The imaginary part of the permeability measured on the 4 films is presented in Fig. 4. The thinnest film displayed a highly resonant permeability, and exhibited a secondary peak at higher frequency. This peak could be attributed to certain inhomogeneities and the excitation of a higher frequency mode.[31] The thicker film exhibited a permeability with a very damped behaviour, and permeability levels down by a factor up to 4. The "efficient dynamic magnetization" $M_\mu(F)$ obtained from the experimental spectra using Eq. (22) is represented in Fig. 5. As expected, this quantity was close to the saturation magnetization at high frequency. Better estimates of the saturation magnetization can be obtained using Eq. (24) with the corrections $s$ and $t$ computed from Eqs (18)-(19). These refined estimates are also shown in the graph, with their associated error bars. It can be seen that all estimate ranges were comprised within the experimental error of the measured $4\pi M_s$. A larger measurement uncertainty for the thin film permeability as evidenced in Fig. 4 was responsible for a larger uncertainty on the integral quantity. This proves that the integral relation on thicker films may be very useful in order to obtain more precise experimental data. It is remarkable that, despite the strong difference in the four spectra presented in Fig. 4, all the estimates derived using Eq. (24) coincided within the experimental errors.

## VI. DISCUSSION AND CONCLUSION

Previous studies[11,12] have established that, for a given set of assumptions, the quantity $\int_0^\infty \mu''(f).f.df$ displays remarkable properties. The present work demonstrates that this is a very general result. In addition, it shows that the finite frequency band on which microwave permeability measurements are performed generally suffices in order to obtain a good estimate of this integral. For materials with magnetic responses significantly influenced by skin effect, one may wonder whether it would not be easier to first determine the intrinsic permeability from measured permeabilities with skin effect, and subsequently determine the integral of the imaginary part of the permeability. Though this procedure is possible, it should be underlined that, for magnetic particles constituted of different layers and/or domains with varying permeabilities, the intrinsic permeability cannot be rigorously obtained from the measurements. In contrast, the estimate of $s$ for the skin effect correction on the integral is independent on the detailed magnetic parameters. It depends only on the $a^2\sigma$ product for the particle, and as a consequence it is a much more robust parameter. In the case of composites made of ferromagnetic powders, there is often a significant size distribution with respect to the particles. As a consequence, the intrinsic permeability as determined by the inversion of Eq. (4) with an averaged value of $a$, may have a limited validity, while the integral quantities can be exploited. In this case, the averaged value of the corrective term $s$ is directly related to $<a^2>$, as expressed by Eq. (15).

The particular properties of $\int_0^F \mu''(f).f.df$ derived in this work may be useful for at least three purposes: microwave measurement, microwave design, and material characterization. Depending on personal likings, one may prefer to work either with the integral, expressed in $Hz^2$; with the dimensionless ratio $k_A$ defined in Eq. (3); or with the efficient dynamic magnetization $M_\mu(F)$ defined by Eq. (22). All properties established for the integral can be easily translated into the two other quantities.

For microwave measurements, the upper bound of the integral (Eq. (21), (28)) can be used to verify the consistency of the measured permeability spectra. This can be done very simply.

It is not necessary for the measurement system to cover a large band since the majoration is valid for any integration range. This is very useful, especially for supporting experimental results claiming large permeability levels, since microwave magnetic measurements are known to be tricky. In the case of thin film permeameters that cover a broad enough bandwidth, the relation $M_\mu(F) \approx 4\pi M_s$ can be used to assess or to demonstrate the measurement precision of the apparatus. In most cases, the uncertainty on the right member is essentially due to the gyromagnetic ratio $\gamma$, which is generally considered to be comprised between 2.8 and 3 GHz/kOe, and to some extent to the measurement precision of the saturation magnetization and of the film thickness.

The results established in this work are useful also for the design of magnetic microwave materials. The sum laws (21), (27), (28) express certain tradeoffs between high permeability levels and operation at high frequencies. As a consequence, these sum laws may be viewed as generalisations of Snoek's law. The integral quantity can easily be experimentally determined on many materials. Simple figures of merit deduced from this integral may be used to compare microwave materials. In some applications, *f.μ''(f)* is a quantity of direct interest. This is the case when magnetic losses are desired for microwave attenuation, either for microwave filtering, electromagnetic compatibility or for Radar Absorbing Materials. The first order approximation of the reflection or transmission losses are in *f.μ''(f)*. It has been shown that in the thin absorber limit, the performance of a magnetic absorber is bounded by the integral.[12] As a consequence, it is an important result that moderate skin effect maintain the integral losses unaffected, even though their frequency distribution is much affected. For somewhat larger skin effect, the correction factor *–s* may become significant. This leads to a decrease in the integrated losses.

Last but not least, this study shows that microwave permeability measurements can be a tool for obtaining information on the magnitude and the orientation of the magnetization within samples. The integral is related to a few magnetic parameters even in cases where *μ''* cannot be described by simple models because of some heterogeneities or magnetic coupling. This is very appealing for the study of unsaturated materials. In the case of multilayers, Eqs. (25) and (26) show that the integral provides an indication on the average orientation of the magnetization within the sample thickness. The microwave field is indeed a probe of the magnetization normal to the excitation, with the ability to penetrate into relatively thick samples. The integral $\int_0^F \mu''(f).f.df$ provides quantitative information on the magnitude of the magnetization normal to the probe field.

## APPENDIX

The dependence of the fields with time is assumed to be exp(+*jωt*), which is consistent with permeabilities that take the form *μ'-jμ''*, *μ''*>0. The expression of the permeability of a uniformly magnetized ellipsoid can be written as:

$$\mu_G = 1 + \frac{F_M(F_y + j\alpha f)}{F_0^2 + j\alpha(F_x + F_y)f - f^2} \quad (A1a)$$

with

$$F_0 = \sqrt{F_x.F_y} \quad (A1b)$$

$$F_M = \varkappa 4\pi M_s \; ; \; F_x = \varkappa(N_x M_s + H_{int}) \; ; \; F_y = \varkappa(N_y M_s + H_{int}) \tag{A1c}$$

$$H_{int} = H_k - N_z M_s \; ; \; \alpha << 1 \tag{A1d}$$

$4\pi M_s$ is the saturation magnetization of the material, $N_x$, $N_y$, $N_z$ are the demagnetizing coefficients of the ellipsoid, and $H_k$ is the external field (or anisotropy field) that saturates the ellipsoid along $+z$. $\varkappa = \gamma/2\pi$ is close to 3 MHz/Oe. $F_0$ is the resonance frequency. The orientation conventions are similar to those in ref [12], the permeability given by (A1a) being in the $x$ direction. In the case of soft magnetic materials under microwave excitation, and for null or moderate external fields, the different contributions to the $H_k$ field will be small as compared to the saturation magnetization. The internal field $H_{int}$ must be positive for the magnetization to be stable in the $+z$ direction, and as a consequence $H_{int}$ is also small.

Following Eq. (6), a central issue is to estimate the quantity

$$\frac{1}{2}\int_0^\pi \mu(Fe^{j\theta}).(Fe^{j\theta})^2 .d\theta$$

for a frequency $F$ much larger than the resonance frequency $F_0$. It is convenient to introduce the reduced frequency

$$\nu = \frac{Fe^{j\theta}}{F_0} \tag{A2}$$

thereby giving

$$\frac{1}{2}\int_0^\pi \mu(Fe^{j\theta}).(Fe^{j\theta})^2 .d\theta = \frac{1}{2} 4\pi\chi_0 F_0^2 \int_0^\pi \frac{\mu(\nu)}{4\pi\chi_0}.\nu^2 .d\theta \tag{A3}$$

The permeability from Eq. (A1) can thus be expressed as a function of the reduced frequency $\nu$:

$$\mu_G = 1 + \frac{4\pi\chi_0}{1-\nu^2 + 2j\beta.\nu}(1 + j\alpha'\nu), \tag{A4a}$$

where $4\pi\chi_0$ is the initial susceptibility,

$$4\pi\chi_0 = \frac{F_M}{F_x} = \frac{F_M.F_y}{F_0^2}; \tag{A4b}$$

$$2\beta = \alpha\frac{F_x + F_y}{F_0} \; ; \; \alpha' = \alpha\frac{F_0}{F_y}. \tag{A4c}$$

This is valid provided that the internal fields are small

$$H_k, H_{int} << 4\pi M_s \tag{A4d}$$

The development in $1/\nu$ of the susceptibility can be written as:

$$\frac{4\pi\chi_G(\nu)}{4\pi\chi_0} = \frac{-1}{\nu^2}\left[1 - \frac{2j\beta}{\nu} - \frac{1}{\nu^2}\right]^{-1}(1 + j\alpha'\nu)$$

$$\approx \frac{-1}{\nu^2}\left(1 + j\alpha'\nu + \frac{2j\beta'}{\nu}\right) \tag{A5}$$

with

$$2\beta' = \alpha \frac{2F_x + F_y}{F_0}. \tag{A6}$$

The terms in $\alpha'.\beta$ have been neglected in the above expression since the damping parameter is small and the terms in $\alpha^2$ are negligible. It should be noted that when the Bloch-Blombergen damping parameter is used instead of the Gilbert damping parameter, the expression of the susceptibility $4\pi\chi_B$ has the same form as Eq. (A1), with $\alpha'=0$ and $\beta=1/T$, where $T$ is the characteristic damping time.

It is thus necessary to obtain an appropriate development of the factor $A(ka)$ that accounts for the skin effect. The wavevector inside the ferromagnetic inclusion can be expressed as:

$$k = \sqrt{\varepsilon.\mu}\,\omega/c, \tag{A7}$$

where $\omega=2\pi f$ is the pulsation corresponding to the frequency $f$, $c$ the celerity of light, and $\varepsilon$ the permittivity of the inclusion.

$$\varepsilon = \frac{-j\sigma}{\omega.\varepsilon_0}, \tag{A8}$$

Here, $\varepsilon_0$ is the dielectric constant of void and $\sigma$ the conductivity. When a skin effect is present but not overwhelming, it is possible to use the low order development of $A(ka)$ according to Eq. (5). In the case where the upper integration frequency $F$ is significantly larger than the gyromagnetic resonance frequency, but not too large,

$$F_0 << F << \frac{1}{2\pi\mu_0 a^2 \sigma}, \tag{A9}$$

we obtain

$$A(\nu) \approx 1 - jb.\nu.(1 + 4\pi\chi_B(\nu)), \tag{A10}$$

with

$$b = \frac{2\pi\mu_0 F_0 a^2 \sigma}{p}. \tag{A11}$$

The set of assumptions (A9) can be written as:

$$1 << |\nu| << 1/(b.p), \tag{A12}$$

The permeability in the presence of skin effect can be written as

$$\mu(\nu) \approx 1 + 4\pi\chi_G(\nu).[1 - 2jb.\nu - jb.\nu.4\pi\chi_B(\nu)] - jb.\nu. \tag{A13}$$

Keeping only the most significant terms leads to

$$\mu(\nu) \approx 1 - \frac{4\pi\chi_0}{\nu^2}.\left[1 + 4b\beta' + jb\left(\frac{4\pi\chi_0}{\nu} - 2\nu\right) + j2\frac{\beta'}{\nu} + j\alpha'\nu\right] - jb.\nu. \tag{A14}$$

The integration of $\mu(\nu).\nu$ on the semi-circle $C^-$ can be expressed as the linear combination of integrals of $\nu^n$ with different powers n. These integrals are easily calculated using:

$$\frac{1}{2}\int_{C^-} \nu^n.d\theta = \frac{1}{2}|\nu|^n \int_0^\pi e^{jn\theta}.d\theta. \tag{A15}$$

One finds

$$\int_0^F \mu''(f).f.df \approx \frac{\pi}{2}(F_M.F_y)\left[1-\frac{2}{\pi}b\left(4\pi\chi_0\frac{F_0}{F}+2\frac{F}{F_0}\right)-\frac{4\beta'}{\pi}\frac{F_0}{F}+\frac{2}{\pi}\alpha'\frac{F}{F_0}\right]+\frac{3}{2}\left(\frac{bF}{F_0}\right).F^2. \quad (A16)$$

Eq. (A16) can be written in the following form

$$\int_0^F \mu''(f).f.df \approx \frac{\pi}{2}N_y(\gamma 4\pi M_s)^2.[1-s-t+g]+S', \quad (A17)$$

where $s$, $S'$, $t$ and $g$ are small corrective terms, with positive signs. $s$ and $S'$ are related to skin effect; $t$ corresponds to the finite truncation; $g$ corresponds to a small contribution when the Gilbert damping model is considered, but becomes zero when the Bloch-Bloembergen damping model is employed. Each of these terms will be discussed in the following sections.

### A. Corrections associated with the magnetic damping

Previous work[11] has been conducted assuming a magnetic damping described by the Bloch-Bloembergen equations. When the damping is described according to the Landau-Lifshitz-Gilbert model, the integral up to infinite frequencies diverges. The corrective terms in Eq. (A17) are:

$$g=\frac{2}{\pi}\alpha'\frac{F}{F_0}=\frac{2}{\pi}\alpha\frac{F}{F_y}=\frac{2}{\pi}\alpha\frac{F}{N_y F_M}. \quad (A18)$$

Cases where $N_y$ is null or small are of no interest since the dominating factor in the expression of the integral is proportional to $N_y$. As the upper integration bound, $F$ is expected to be lower than $F_M$, and since $\alpha$ is small (from a few percent down to a fraction of a percent), $g \ll 1$. This establishes that for a practical case, the theoretical results obtained for $\int_0^F \mu''(f).f.df$ are independent of the damping model under consideration.

### B. Skin effect corrections

Let us examine in more detail the corrective terms associated with skin effect in Eq. (A17). The term $S'$ is independent of the magnetization of the sample.

$$S'=\frac{3}{2}\left(\frac{bF}{F_0}\right).F^2. \quad (A19)$$

It is well known that conductive particles may exhibit non-unit permeability as a result of eddy currents. Although the integral of $\mu''(f).f$ diverges at infinity, it should be noted that if the integral is performed only up to a frequency $F$ that is not too large, then $S' \ll F^2$ according to Eq. (A9). In the case of ferromagnetic materials, it is convenient to express $S'$ as a perturbation of the main term in Eq. (A17):

$$S'=\frac{\pi}{2}N_y F_M^2.s'. \quad (A20)$$

The expression for $s'$ is:

$$s'=\frac{3}{\pi N_y}\left(\frac{bF}{F_0}\right).\left(\frac{F}{F_M}\right)^2=\frac{6\mu_0 a^2 \sigma}{p}.\frac{F^3}{F_M^2}. \quad (A21)$$

and $s'$ is negligible provided that (A9) is met and $F<F_M$. Let us now examine the $s$ corrective term in Eq. (A17),

$$s=\frac{2}{\pi}b\left(4\pi\chi_0\frac{F_0}{F}+2\frac{F}{F_0}\right)\approx\frac{4\mu_0 a^2\sigma}{p}N_y\frac{F_M^2}{F}. \quad (A22)$$

It is straightforward that *s* is positive, which means that skin effect tends to decrease the value of the integral.

### C. Finite frequency summation correction

Stopping the integration at a certain finite upper frequency would affect the value of the integral, even if no skin effect is present. The *t* term in Eq. (A17) accounts for this truncation effect. It can be expressed as:

$$t = \frac{4\beta'}{\pi}\frac{F_0}{F} \approx \frac{2}{\pi}\alpha(2N_x + N_y)\frac{F_M}{F}. \tag{A23}$$

This correction factor is far below unity provided that

$$F >> \alpha F_M = \alpha(\gamma 4\pi M_s). \tag{A24}$$

Even for a material with a very large saturation magnetization such as CoFe with $4\pi M_s$=24k Oe, for a typical value of $\alpha$=2%, Eq. (A24) requires that the upper integration frequency is such that *F*>>1.4 GHz. This is an easily met condition.

### D. Effect of experimental measurement errors

When using experimental permeability data, the error on the sum increases when the upper integration bound is extended. An error term has to be added to the right member of Eq. (A17). In order to be able to directly compare the error $\pm e$ to the other terms –*s* and –*t*, it is convenient to write this additional term as

$$\frac{\pi}{2}N_y F_M^2.e.$$

It can thus be shown in a straightforward manner that

$$|e| \leq \frac{\Delta\mu}{\pi N_y}\left(\frac{F}{F_M}\right)^2, \tag{A25}$$

where $\Delta\mu$ is the maximum error on the measured permeability. The relative error is small provided that

$$F << (\gamma 4\pi M_s)/\sqrt{\Delta\mu}. \tag{A26}$$

**Figure Caption**

| inclusion type | sketch | $A(ka)$ | 2nd order approx. |
|---|---|---|---|
| plate | 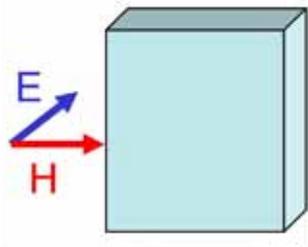 | $\dfrac{\tan(ka)}{ka}$ | $1 + \dfrac{(ka)^2}{3}$ |
| cylinder $\perp H$ | 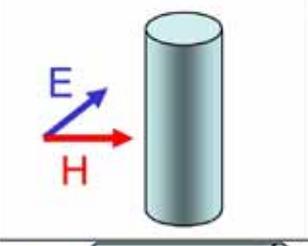 | $\dfrac{J_1(ka)}{ka \cdot J_0(ka) - J_1(ka)}$ | $1 + \dfrac{(ka)^2}{4}$ |
| cylinder $// H$ | 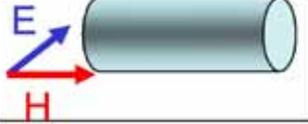 | $\dfrac{2 J_1(ka)}{ka \cdot J_0(ka)}$ | $1 + \dfrac{(ka)^2}{8}$ |
| sphere | 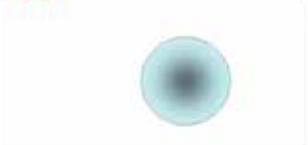 | $\dfrac{2(\tan(ka) - ka)}{ka + ((ka)^2 - 1)\tan(ka)}$ | $1 + \dfrac{(ka)^2}{10}$ |

FIG. 1. Various inclusion shapes, with the associated functions *A(ka)* used for the expression of the permeability in the presence of skin effect according to ref [19]; *k* is the wavevector inside the inclusion, and *a* is its radius (or half thickness in the case of a plate). The expression of the 2nd order approximation of *A(ka)* is also given.

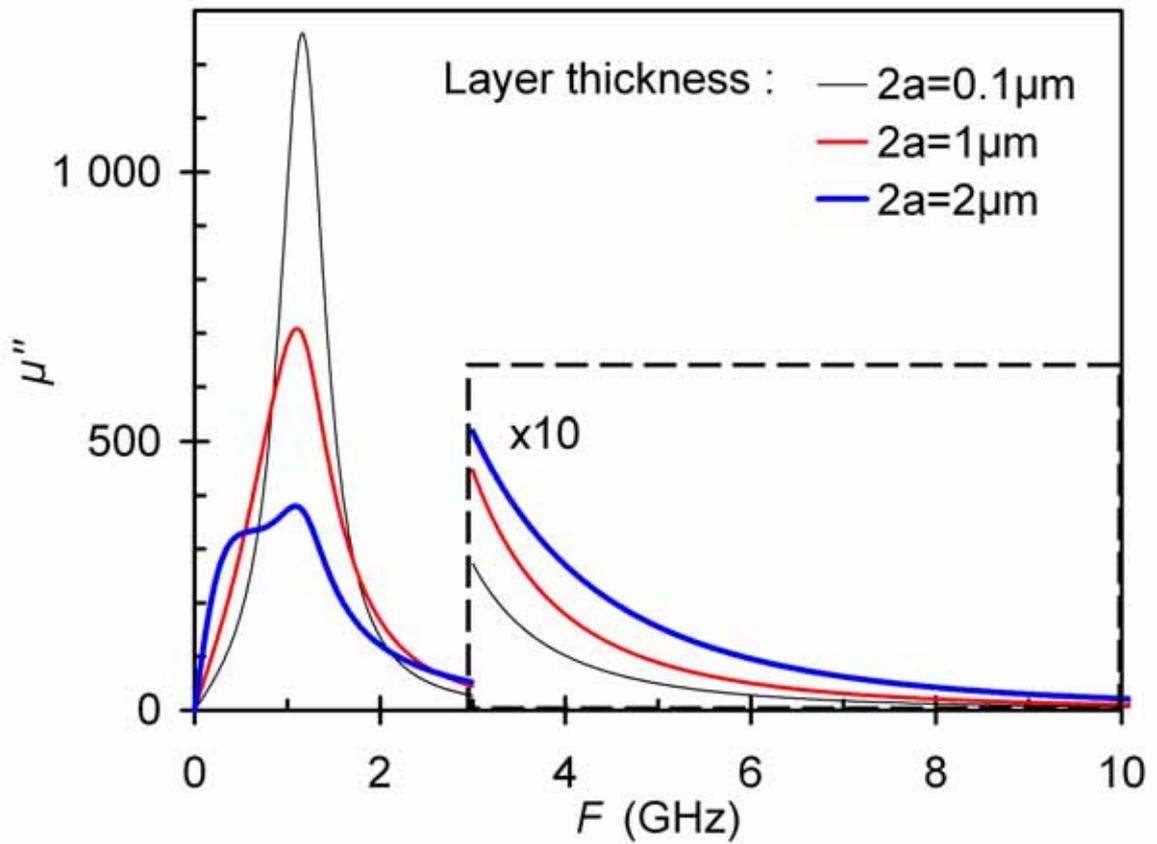

FIG. 2 The imaginary part of the permeability $\mu''$ computed for films with a thickness $2a$ ranging from 0.1 µm to 2 µm, using the Landau-Lifschitz-Gilbert model and taking into account skin effect.

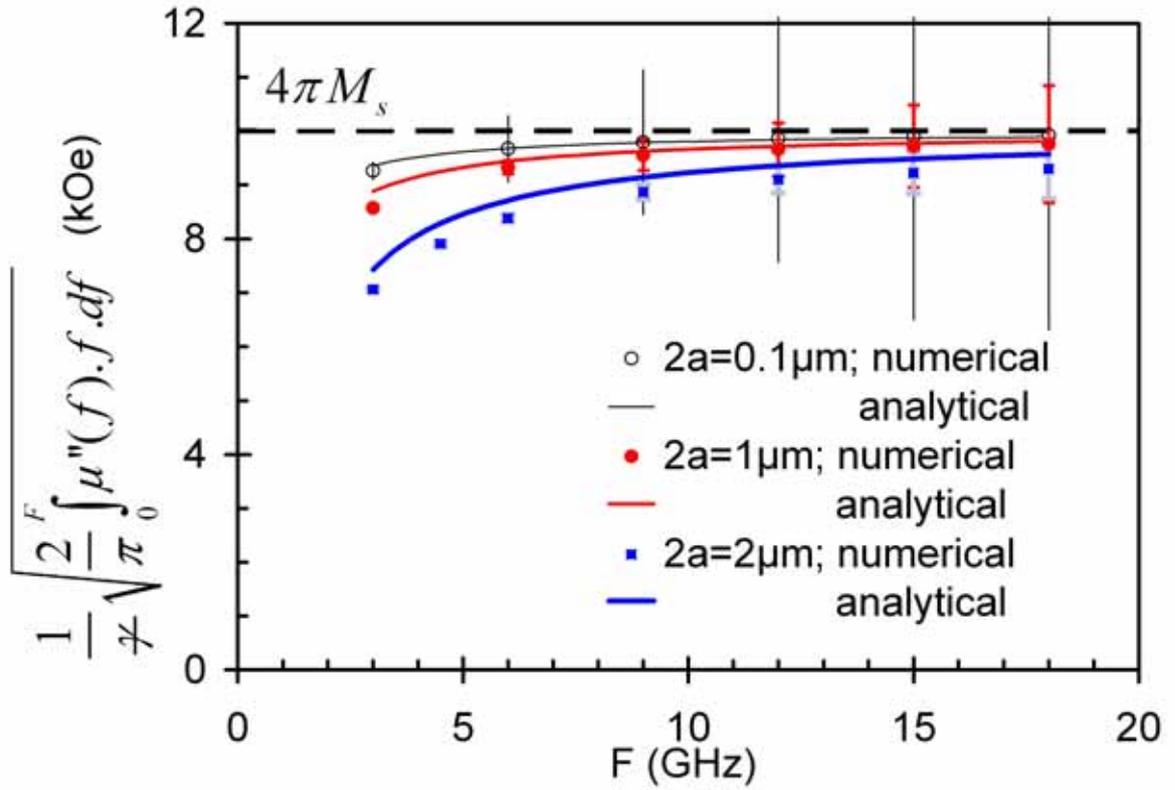

FIG 3. The efficient dynamic magnetization $M_\mu(F)$ associated with the calculated permeabilities represented in Fig. 2, obtained either by numerical integration (symbols), or by analytical estimates (lines). The error bars represent typical experimental errors and the dashed line corresponds to the saturation magnetization.

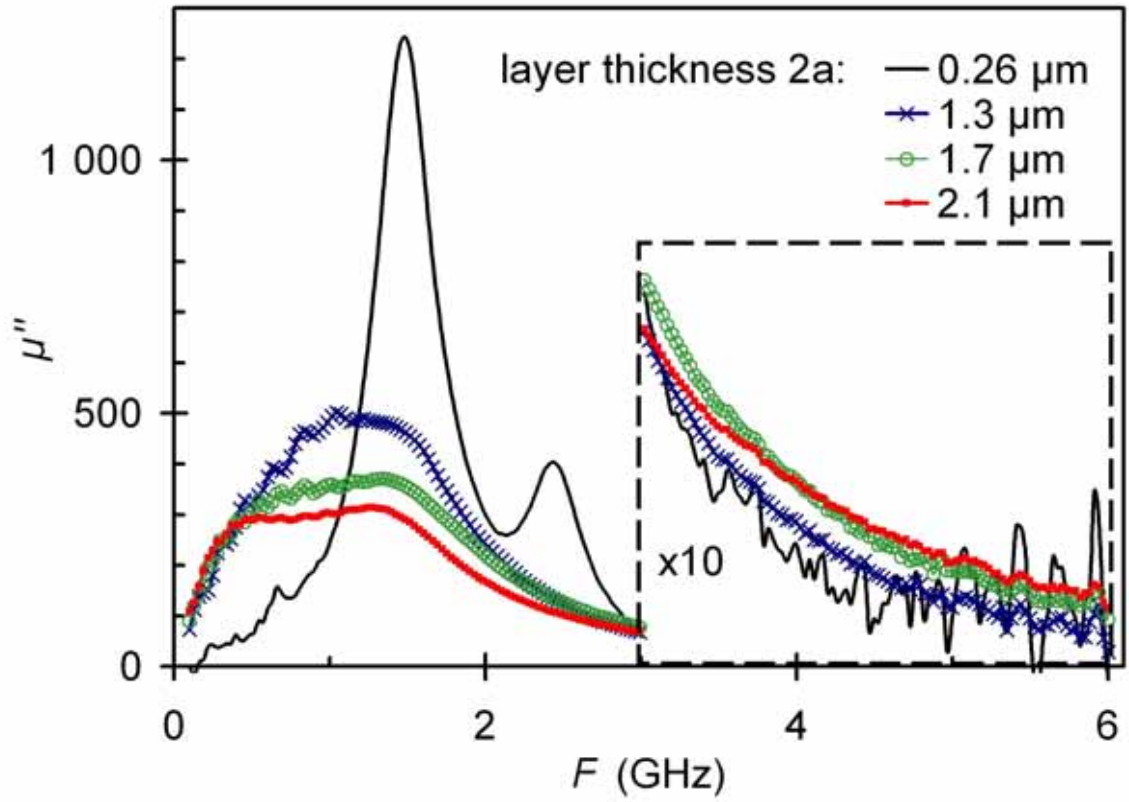

FIG. 4. The imaginary part of the permeability μ'' measured on CoZr amorphous thin films of varying thicknesses.

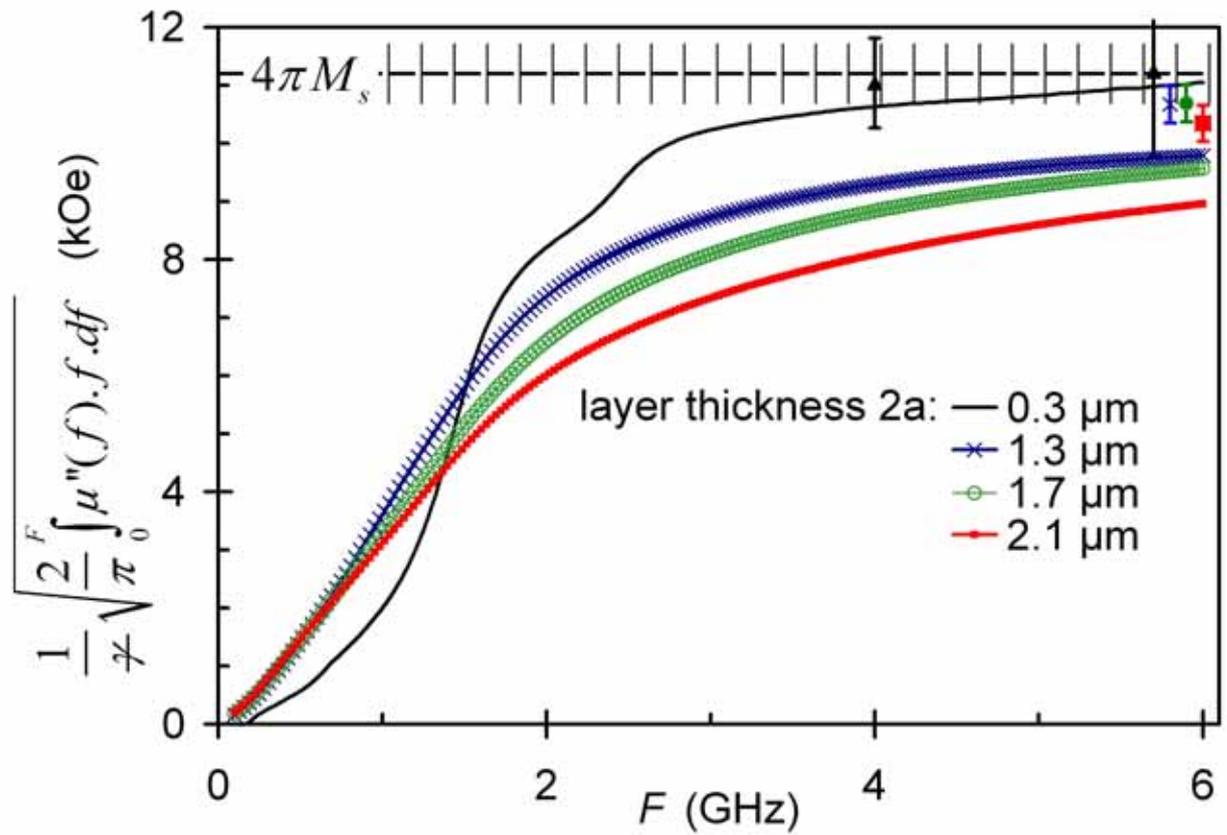

FIG. 5. The efficient dynamic magnetization $M_\mu(F)$ associated with the measured permeabilities represented in Fig. 4, and a comparison with the saturation magnetization values (dashed line with error bars). The extrapolation of the saturation magnetization from $M_\mu(F)$ using analytical estimates is also represented, with appropriate error bars.

Table Caption

| Lower bound | Upper bound | Eq. | Comment |
|---|---|---|---|
| $F_0 << F$ | $F << (2\pi\mu_0 a^2 \sigma)^{-1}$ | (A9) | Required for the validity of the whole approach |
| | | (A4d) | $H_k, H_{int} << 4\pi M_s$ also required |
| | $F << F_M/\alpha$ | (A18) | Effect of Gilbert damping $+g$ is small |
| | $F < F_M$ | (A21) | Skin effect term $s'$ is small |
| $4\mu_0 a^2 \sigma F_M^2 / p << F$ | | (A22) | Skin effect term $-s$ is small |
| $\alpha.F_M << F$ | | (A24) | Truncation term $-t$ is small |
| | $F << F_M/\sqrt{\Delta\mu}$ | (A25) | Experimental error $\pm e$ is small |

Table I. Guidelines for choosing the upper integration frequency F on $\int_0^F \mu''(f).f.df$, as a function of the gyromagnetic resonance frequency $F_0$, $F_M = \gamma 4\pi M_s$, and other parameters.

| Lower bound | Upper bound | Parameter | Eq. | Analytical Value | Numerical value |
|---|---|---|---|---|---|
| $F_0$=1.2 GHz | $(2\pi\mu_0 a^2 \sigma)^{-1} = 165$ GHz | $\int_0^F \mu''(f).f.df$ | | $1.05\ 10^3$ GHz$^2$ | $0.99\ 10^3$ GHz$^2$ |
| | $F_M/\alpha$=1500 GHz | $+g$ | (A18) | 0.3% | 0.4% |
| | $F_M$=30 GHz | $+s'$ | (A21) | 0.05% | <<0.1% |
| $\frac{4\mu_0 a^2 \sigma F_M^2}{p} = 1.2$ GHz | | $-s$ | (19) | -19% | -23% |
| $\alpha.F_M$=0.6 GHz | | $-t$ | (18) | -6.4% | -6.5% |
| | $F_M/\sqrt{\Delta\mu} = 30$ GHz | $\pm e$ | (20) | ±1.3% | |

Table II. The numerical estimation of the lower and upper frequency bounds for $F$ associated with Table I in the case of a 2 µm thick film; for $F$=6 GHz, values of the different corrective terms accounting for the relative difference between $\int_0^F \mu''(f).f.df$ and $\frac{\pi}{2}(\gamma 4\pi M_s)^2$ obtained both analytically and numerically.